\documentclass{article}
\usepackage{graphicx}
\usepackage{amsmath}
\usepackage{a4}

\newtheorem{theorem}{Theorem}

\newenvironment{proof}[1][Proof]{\textbf{#1.} }{\ \rule{0.5em}{0.5em}}

\begin{document}

\title{Dressing Chain Equations Associated to Difference Soliton Systems }
\author{S. Leble \\
Technical University of Gda\'{n}sk, ul. Narutowicza 11/12\\
80-952 Gda\'{n}sk , Poland\\
Kaliningrad State University, ul. A. Nevskogo, 14\\
236041 Kaliningrad, Russia.}
\maketitle

\begin{abstract}
Links of factorization theory, supersymmetry and Darboux transformations as
isospectral deformations of difference operators are considered in the context
of soliton theory.
The dressing chain equations for factorizing operators of a spectral problem
are derived. The chain equations itselves yield nonlinear systems which
closure generates solutions of the equations as well as of the nonlinear
system if both operators of the correspondent Hirota bilinearization are
covariant with respect to Darboux transformation which hence defines a
symmetry of the nonlinear system as well as of these closed chains.
Examples of Hirota and Nahm equations are specified.
\end{abstract}
\section{ Introduction.}

The growing interest to discrete models  appeal to necessity to
widen classes of  symmetry structures of the correspondent
nonlinear problems \cite{Ne}. Such theories as conformal field
result  in set of equations to be investigated \cite{Bel}. Next
kind of problems contain solvable lattice models \cite{Mi},
\cite{Kun}.

Very recently a good basis for new searches in this field of
differential-difference and difference-difference equations was
discovered \cite{Mat} in the context of classical Darboux
Transformations (DT) theory development. Likely the differential
operator case it would have links to Hirota bilinearization method
\cite{LLS} and to factorization theory \cite{LeZa} with similar
applications possibilities.

One of very promising approach to the solutions construction is
based on the dressing chain notion \cite{IH}, \cite{W}, \cite{SH},
\cite{VS}.  The approach cover soliton, rational, finite-gap and
other new solutions inside universal scheme, reducing the problem
to a solution of closed sets of nonlinear ordinary equations with
bi-Hamiltonian structure \cite{VS}.

In this paper we reformulate Darboux covariance theorem from
\cite{Mat} introducing a kind of the difference Bell polynomials.
Such polynomials have natural correspondence  to the differential
(generalized) Bell polynomials  in its nonabelian version
\cite{LeZa}, \cite{Faa}.
 Its usage shortens the transformation formulas and helps to apply the theory
 in complicated cases of joint covariance of  U-V pairs  \cite{L}, \cite{LLS}, \cite{Le} on
 a way of the dressing chain equations derivation.

 The example of Hirota equation \cite{H}, connected with known applications, is studied in
 \cite{Mat}. Here we derive the dressing chain equations and study simplest solutions of
 it. In the last section we propose the lattice Lax pair for Nahm equations \cite{Nahm},
 see also \cite{HHM}, \cite{Ne} on physical aspects. The Lax pair is covariant with
  respect
 to a combined Darboux-gauge transformations that generates the dressing structure for the
 equations.

\section{ Darboux transformations in associative ring\newline
with automorphism.}

In this section we reformulate, and analyze the results from \cite{Mat}
 in a context of its further use in the chain equations derivation and joint
  covariance of operators pairs \cite{L} investigation.
We would begin with general notations. Let \textbf{R} be an
associative ring with an automorphism,  implying that there exist
a linear infertile map $T,$ \textbf{R}$\rightarrow $\textbf{R}\
such that for any $\psi\left( x,t\right) ,\varphi\left( x,t\right)
\in $\textbf{R, }$x\in R^{n},t\in \bf{R},$

\begin{equation}
T(\psi\varphi)=T(\psi)T(\varphi),\;T(1)=1.  \label{aut}
\end{equation}

The automorphism with the defining property $\left( \ref{aut}\right) $
allows to write down a wide class of functional-differential-difference and
difference-difference equations \cite{Mat} starting from
\begin{equation}
\psi_{t}\left( x,t\right) =\sum_{m=-M}^{N}\ U_{m}T^{m}\psi.
\label{eq1}
\end{equation}

For example, the operators T could be chosen as,
\begin{equation}
T\psi\left( x,t\right) =\psi\left( qx+\delta ,t\right) ,
\label{T}
\end{equation}
where $q\in GL(n,C),\delta \in R^{n}.$

\begin{equation*}
T\psi\left( x \right) =W\psi(x)W^{-1}\ ,W\in GL(n,C)
\end{equation*}
We would save the notations and conditions of the mentioned paper
\cite{Mat} discussing other potentials definitions in the end of
the section 4.

Let us consider the definitions of two Darboux transformations for
solutions of \eqref{eq1},
\begin{equation}
D^{\pm }f=f-\sigma ^{\pm }T^{\pm 1}f,\;\sigma ^{\pm }=\varphi
\left( T^{\pm 1}\varphi \right) ^{-1}.  \label{DT}
\end{equation}
where $\varphi$ is a particular solution of the same equation
$\left( \ref{eq1}\right) $
For the case of a differential ring and for  $%
Tf\left( x,t\right) =f\left( x+\delta ,t\right) ;x,\delta \in R$
the limit $\partial f=\lim_{\delta \rightarrow 0}\frac{1}{\delta
}(T-1)f\left( x,t\right)$   give the link to the classical DT.

To derive the DT of potentials $U_{m}$ it is necessary to evaluate
the derivative of the elements $\sigma ^{\pm }$ with respect to
the variable t (say - time). We shall do it by means of the
special functions (differential Bell polynomials analogue)
introduction similar to \cite{LeZa} . Let us start
from the first version of the DT definition $D^{+},$ expressing
the $T\varphi $ from $\left( \ref{DT}%
\right) ,$ hence $T\varphi =\left( \sigma ^{+}\right) ^{-1}\varphi .$ Acting
to the result by $T$, taking into account $\left( \ref{aut}\right) $ yields $%
T^{2}\varphi =T\left( \left( \sigma ^{+}\right) ^{-1}\right) T\varphi
=T\left( \left( \sigma ^{+}\right) ^{-1}\right) \left( \sigma ^{+}\right)
^{-1}\varphi .$ Repeating the action, one arrives at

\begin{equation}
T^{m}\varphi =\prod\limits_{k=0}^{m-1}(T^{k}\left( \sigma ^{+}\right))^{-1} \varphi
=B_{m}^{+}\left( \sigma ^{+}\right) \varphi .  \label{BP}
\end{equation}
Here and below the product is ordered by the index ($k$) from
right to left.

{\bf Definition} The equation \eqref{BP} defines the function
$$B_{m}^{+}\left( \sigma \right) =
\prod\limits_{k=0}^{m-1}T^{k}\left( \sigma \right)^{-1})$$

It is convenient to write down the t-derivative of the $\sigma $
by means of the functions $B_{m}^{+}\left( \sigma ^{+}\right) $ \
that are connected generalized with Bell polynomials \cite{LeZa}.
\begin{equation}
\sigma _{t}^{+}=\sum\limits_{m=-M}^{N}\left[ U_{m\ }B_{m}^{+}\left( \sigma
^{+}\right) \sigma ^{+}-\sigma ^{+}T\left( U_{m\ }\right) B_{m+1}^{+}\left(
\sigma ^{+}\right) \sigma ^{+}\right]  \label{Mi}
\end{equation}

The resulting equation $\left( \ref{Mi}\right) $ is some nonlinear equation
associated with the $\left( \ref{eq1}\right) $ that is reduced to a
generalized Miura transformation in the stationary case (see sec.3).

The Matveev theorem \cite{Mat} gives powerful generalizations of
the conventional Darboux theorem, proved originally  for the
second order differential equation (for generalizations look
\cite{MS} as well) and can be formulated by means of the
introduced entries in the following way.

\begin{theorem}
Let the functions $\varphi \in $\textbf{R} and $\psi$ $\in
$\textbf{R }satisfy
the equation $\left( \ref{eq1}\right) $, then the \ function $\psi^{+}=$ $%
D^{+}\psi $ satisfies the equation $\psi_{t}^{+}\left( x,t\right)
=\sum_{m=-M}^{N}\ U_{m}^{+}T^{m}\psi^{+},$ where the coefficients
are evaluated from the following recurrent relations
\begin{equation}
U_{-M}^{+}=U_{-M}  \label{U+}
\end{equation}
\begin{equation}
U_{1}^{+}-U_{0}^{+}\sigma ^{+}=U_{1}-\sigma ^{+}TU_{0}-\sigma^+_{t}
\label{U1+}
\end{equation}
\begin{equation}
U_{m}^{+}-U_{m-1}^{+}T^{m-1}\sigma ^{+}=U_{m}-\sigma ^{+}TU_{m-1}
\label{Um+}
\end{equation}
\begin{equation}
U_{N}^{+}\ =\sigma ^{+}\left( TU_{N}\right) \left( T^{N}\sigma ^{+}\right)
^{-1}  \label{UN+}
\end{equation}
The expressions $\left( \ref{U+}\right) -$ $\left( \ref{UN+}\right) $ define
the Darboux transformations of the coefficients of the differential equation $%
\left( \ref{eq1}\right) $ (potentials) as the recurrence. The
straight corollary of this statement solving the recurrence
\eqref{Um+} by means of $\left( \ref{BP}\right) $ yields
\begin{equation}
U_{m}^{+}=\sum_{l=0}^{m+M}U_{-M+l}-\sigma ^{+} \left(
TU_{-M+l-1}\right) B_{-M+l}^{+}\left( \sigma ^{+}\right) \left(
B_{m}^{+}\left( \sigma ^{+}\right) \right) ^{-1}  \label{DDT2}
\end{equation}
\begin{equation}
U_{N}^{+}\ =\sigma ^{+}\left( TU_{N}\right) \left( T^{N}\sigma
^{+}\right) ^{-1}  \label{DDT3}
\end{equation}

\end{theorem}
The theorem establishes the covariance (form-invariance) of the equation $%
\left( \ref{eq1}\right) $ with respect to the Darboux transformation.

\begin{proof}
For the proof  it is necessary to check the additional equality
that appears by the term $T^{m}\psi$ with essential use of the
expression for $\sigma _{t}^{+}$ from \eqref{Mi}.
\end{proof}

The formalism for the second DT from \ $\left( \ref{DT}\right) $
may be constructed similarly on the base of the identity
\begin{equation}
T^{m}\varphi =\prod\limits_{k=0}^{m}T^{k}\left( \sigma ^{-}\right)
T^{-1}\varphi =B_{m}^{-}\left( \sigma ^{-}\right) T^{-1}\varphi.
\label{BP1}
\end{equation}
The definition of the second type lattice Bell polynomials
$B_{m}^{-}\left( \sigma ^{-}\right)$ may be extracted from
\eqref{BP1}.
The evolution equation for $\sigma^{-} $\ is similar to $\left( \ref{Mi}\right) $%
\begin{equation}
\sigma _{t}^{-}=\sum\limits_{m=-M}^{N}\left[ U_{m\ }B_{m}^{-}\left( \sigma
^{-}\right) -\sigma ^{-}T^{-1}\left( U_{m\ }\right) B_{m-1}^{-}\left( \sigma
^{-}\right) \right]  \label{Mi1}
\end{equation}
and gives the second generalized Miura map for a stationary solutions of the $%
\left( \ref{eq1}\right).$ Explicit formulas for $U_m^-$ are
similar to \eqref{U1+}, \eqref{Um+}, \eqref{UN+}, \eqref{DDT2},
\eqref{DDT3}.

\section{\protect\bigskip\ Stationary equations as eigenvalue problems and
chains}

\bigskip The stationary equation that corresponds to (\ref{eq1}) appears
when the solutions of the following constraint equations
$\psi_{t}\left( x,t\right) =\lambda \psi,\;\psi_{t}\left(
x,t\right) = \psi\lambda ,$ are considered, or, in the first case,
\begin{equation}
\sum_{m=-M}^{N}\ U_{m}T^{m}\psi = \lambda \psi .  \label{sp0}
\end{equation}

The derivative
 \begin{equation}
\sigma_t = \varphi_t(T\varphi)^{-1} -
\varphi(T\varphi)^{-1}(T\varphi)_t(T\varphi)^{-1} = \mu \sigma -
\sigma \mu .  \label{sigt}
\end{equation}
It is zero if the $\sigma $ and $\mu $ commute. Recall that
$\varphi \in \{\psi_{\lambda=\mu}\}.$

One arrives at the algebraic connection between potentials and $\sigma $
that is the corollary of either $\left( \ref{Mi}\right) $ or $\left( \ref
{Mi1}\right) $ in this case of $\left( \ref{sigt}\right) $. When the reduction is such as all
the coefficients in (\ref{eq1}) are functions of some unique potential u,
the connection, allows in principle to express the potential u as the
function of $\sigma $. This connection has the same form as for the potential
u as for a ''dressed'' one. If one takes connections for both ones with the
correspondent sigmas and plug the result into the Darboux
transformation formulas, the chain equations yield.

Let us illustrate the algorithm by an

{\it Example of generalized ZS problem. }

Let us take the equation $\left( \ref{eq1}\right) $ in the minimal variant

\begin{equation}
\psi_{t}\left( x,t\right) =\ \ \left( U_{0}+U_{1}T\right)\psi.
\label{ZS}
\end{equation}
The covariance of this equation with respect to the DT of first kind means
the invariance of $U_{0}.$ We rewrite the transform of $U_{1}$ by means of  $%
\left( \ref{UN+}\right) $
\begin{equation}
U_{1}^{+}=\sigma ^{+}\left( TU_{1}\right) \left( T\sigma ^{+}\right) ^{-1}
\label{DTZ}
\end{equation}
For the spectral problem
\begin{equation}
\ \left( U_{0}+U_{1}T\right) \varphi =\mu \varphi  \label{ZSP}
\end{equation}
the connection of the unique potential $U_{1}$ and $\sigma ^{+}$ is obtained
form $\left( \ref{Mi}\right) $ and $\left( \ref{sigt}\right) $ or, directly
from the equation $\left( \ref{ZSP}\right) ,$

\begin{equation}
\ U_{1}=(\mu -U_{0})\sigma ^{+}\   \label{S+}
\end{equation}

Introducing the number of iteration n for $U_{1},\ $ and n+1 for $U_{1}^{+}$
yields

\begin{equation}
U_{1}\left[ n+1\right] =(\mu -U_{0})\sigma ^{+}[n+1]=\sigma ^{+}\left[ n%
\right] T\left( (\mu -U_{0})\sigma ^{+}\ \left[ n\right] \right)
\left( T\sigma ^{+}\left[ n\right] \right) ^{-1}.  \label{CHZS}
\end{equation}

It could be the chain equation for the generalized ZS problem. We
rewrite the chain equation $\left( \ref{CHZS}\right) $ in more
compact form changing the notations as follows $U_{0}\rightarrow
J,\; U_{1}\rightarrow U, \; \sigma ^{+}\left[ n\right] \rightarrow
\sigma _{n}$ and supposing that $T(J)=J$.  We arrive at

\begin{equation}
\sigma _{n+1}\ \ =\ (\mu -J)^{-1}\sigma _{n}\ (\mu -J)
\end{equation}
 This dressing however is obviously almost trivial.
Such phenomenon is well-known in the differential operator case
\cite{MS}. The alternative and effective transformations appear if
in the stationary equation \eqref{ZSP} one introduces the element
$\mu$ that does not commute with $\sigma$ and changes the order in
the r.h.s. between the elements $\mu$ and $\varphi$. (It is the
second case mentioned at the very beginning of this section.)
\begin{equation}
\ \left( J+U T\right) \varphi =\varphi \mu. \label{ZSP2}
 \end{equation}
 The formula for the potential is changed
\begin{equation} \ U =\varphi \mu (T\varphi)^{-1} - J\sigma
\   \label{S+}
\end{equation}
 and the role of $\sigma_+$ can
play the function $s = \varphi \mu (T\varphi)^{-1}$. The equation
\eqref{Mi} connects U and $\sigma$
\begin{equation}
 J\sigma + U -
\sigma J \sigma + \sigma T(U)(T\sigma)^{-1}=[\mu,\sigma].
\label{S+U}
 \end{equation}
The algorithm of the explicit derivation of the chain equations
begins from the equation (\eqref{S+U}) solving with respect to U
in appropriate way. For matrix rings, it may be a system of
equations for matrix elements, that could be effective
 in low matrix dimensions of (\eqref{S+U}),
as in \cite{SH}. Otherwise it opens the special problem. Let us
rewrite (\ref{S+U}) and the DT \ref{CHZS} in terms of $s$,
excluding U from \ref{S+}, denoting the number of iterations by
index
$$
U[n] = s_n -J\sigma_n.
$$
The equation  (\ref{S+U}) transforms as
\begin{equation}
 s -
\sigma J \sigma + \sigma T(s)(T\sigma)^{-1} + \sigma J =
[\mu,\sigma]. \label{S}
 \end{equation}
 The use of this result give for the DT
\begin{equation}
 s_{n+1} - s_n = J \sigma_{n+1} + \sigma_n J \sigma_n - [\mu,\sigma_n].
\label{S}
 \end{equation}

Then, taking the result (\ref{S})for two indices (e.g. it could be
n,n+1) one have the chain system. In the section 5 we will give
the explicit example for the bilinear Hirota equation.

Let us mention that the chain equations for the classical ZS
problem and two types of DT transformation were introduced in
\cite{SH}. The closure of the chain equations specify classes of
solutions. The chain equations for a standard Sturm-Liouville
problem were derived in connection with quantum mechanical
problems in early Schr\"odinger paper (look the review \cite{IH}).
In connection with the celebrated scalar KdV  important
possibility is studied in \cite{W}. A periodic closure of the
chains produces integrable bihamiltonian finite-dimensional
systems and, in some special cases, the finite-gap potentials
\cite{VS}.

\section{Joint covariance of equations and nonlinear problems}

Let us consider a pair equations of the same type $\left( \ref{eq1}%
\right) $ for a function $\psi $

\begin{equation}
\psi _{t}\left( x,t\right) =\sum_{m=-M}^{N}\ U_{m}T^{m}\psi ,  \label{para1}
\end{equation}

\begin{equation}
\psi _{y}\left( x,t\right) =\sum_{m=-M^{\prime }}^{N^{\prime }}\
V_{m}T^{m}\psi .  \label{para2}
\end{equation}

The compatibility condition for them is the nonlinear equation

\begin{equation}
U_{sy}-V_{st}=\sum_{k}V_{k}T^{k}\left( U_{s-k}\right) -U_{s-k}T^{s-k}\left(
V_{k}\right)  \label{NE}
\end{equation}

for $s=-M-M^{\prime },...,N+N^{\prime },\;k\in \{k^{\prime }=\ -M^{\prime
},...,\ N^{\prime }\}\cap \{s-k=-M\ ,...,N\ \}.$

In the simplest case of ZS operators in both $\left( \ref{para1}\right)
,\left( \ref{para2}\right) $ with the subclass of stationary in y solutions
one have three conditions
\begin{equation}
U_{0t}\ =\ V_{0}U_{0}-U_{0}V_{0},  \label{eqU0}
\end{equation}
\begin{equation}
U_{1t}\ =\ V_{0}U_{1}-U_{0}V_{1}+V_{1}T\left( U_{0}\right) -U_{1}T\left(
V_{0}\right)  \label{eqU1}
\end{equation}

\begin{equation}
V_{1}T\left( U_{1}\right) =U_{1}T\left( U_{1}\right)  \label{eqU3}
\end{equation}

The connection with polynomials of a differential operator and hence with classic
Bell polynomials theory may be found if one change the definition of
potentials. It is clear that if the automorphism T is the shift operator $%
Tf(x)=f(x+\delta ),$
the coefficients of the polynomials in T should be arranged as follows.
\begin{equation}
\psi_{t}\left( x,t\right) =\sum_{m=-M}^{N}\ u_{m}\frac{1}{\delta
^{m}} \ \sum_{r=0}^{m}\binom{m}{m-r}(-1)^{m-r}T^{r}\psi.
\label{T-D}
\end{equation}
The Newton binomial formula have been used. The recursion equation
that define classic differential Bell polynomials \cite{Faa}
\begin{equation}
B_{m+1} = \sum_{r=0}^{m}\binom{m}{r}B_{m-r}y_{r+1}, \label{B}
\end{equation}
together with the definition \eqref{BP} of $B_m^+$ connects the
expression for these special functions. Let us remark that the
transformations for $U_m$ found in the sec. 2 give the transforms
for $u_m$ defined by \eqref{T-D}. The possibility of inverse
transition depends on the independence of functions $(T-1)^nf$ for
a given T and the set of functions $\psi$ under consideration. The
joint covariance of the system \eqref{para1}, \eqref{para2} hence
may be investigated along the guidelines of \cite{LLS}, \cite{Le},
where so-called Binary Bell polynomials are used to form a
convenient basis.

\section{Nonabelian Hirota system }
\subsection{Derivation of the chain equation}
Let us consider a pair of Zakharov-Shabat type equations,

\begin{equation}
\psi_{t}\left( x,y,t\right) =\ \ \left( V_{0}+V_{1}T\right) \psi,
\label{LT1}
\end{equation}

\begin{equation}
\psi_{y}\left( x,y,t\right) =\ \ \left( U_{0}+U_{-1}T^{-1}\right)
\psi. \label{LT2}
\end{equation}
It differs from one used in previous section by the change $T
\rightarrow T^{-1}$ in the r.h.s.


In a t-lattice version of the equation $\left( \ref{eq1}\right) $ with $%
j\in Z,$ one goes to

\begin{equation}
f\left( x,j+1\right) =\sum_{m=-M}^{N}\ U_{m}T^{m}f.  \label{DE}
\end{equation}

The lattice in all variables case is generated by the transition to the discrete variables $%
x,y,t\rightarrow n,j,r\in Z;\;f\left( x,y,t\right) \rightarrow f_{n}\left(
j,r\right) ,$ defined as in \cite{Mat}. The operator $T$ acts as the shift
of n:\ $Tf_{n}\left( j,r\right) =f_{n+1}\left( j,r\right) .$ The corresponding equations
of the system $\left( \ref{LT1}\right) ,\left( \ref{LT2}\right) $ looks as
\begin{equation}
f_{n}\left( j-1,r\right) =f_{n+1}\left( j,r\ \right) +v(n,j,r)f_{n\ }\left(
j,r\ \right)  \label{LH1}
\end{equation}
\begin{equation}
f_{n}\left( j,r-1\right) =f_{n}\left( j,r\ \right) +u(n,j,r)f_{n-1}\left(
j,r\ \right)  \label{LH2}
\end{equation}
with a specified potentials. The compatibility condition of the
linear equations $\left( \ref{LH1}\right) ,\left( \ref{LH2}\right)
$ has the form
$$
u(n,j-1,r)-u(n+1,j,r) = v(n,j,r-1)-v(n,j,r),
$$
\begin{equation}
v(n,j,r-1) u(n,j,r) = u(n,j-1,r) v(n-1,j,r). \label{HB2}
\end{equation}
The second equation of (\ref{HB2}) is automatically valid if
$$u(n,j,r) =  \tau _{n+1}\left( j,r-1\ \right)\tau _{n}^{-1}\left(
j,r -1\right)\tau _{n-1}\left( j,r\ \right)  \tau _{n}^{-1}\left(
j,r\ \right)$$
 \begin{equation}
v(n,j,r) = \tau _{n+1} \left( j-1,r\ \ \right)
\tau_{n}^{-1}\left(j-1,r \right) \tau _{n}  \left( j,r\ \right)
\tau _{n+1}^{-1}\left( j,r\ \right)  \label{SUB}.
\end{equation}
We stress that the form of these expressions is valid for the
order of entries that had been shown in \eqref{HB2}, \eqref{SUB}:
the form differs (generalize) from \cite{Mat}. The substitution of
the (\ref{SUB}) to the first equation from (\ref{HB2}) leads to
the generalized Hirota bilinear equation \cite{H}, compare also
with generalizations in \cite{Mi}
$$
\tau _{n+1}\left( j-1,r-1\ \right)\tau _{n}^{-1}\left( j-1,r
-1\right)\tau _{n-1}\left( j-1,r\ \right)  \tau _{n}^{-1}\left(
j-1,r\ \right) - $$
$$
 \tau _{n+1} \left( j-1,r-1\ \ \right)) \tau
_{n}^{-1}\left( j-1,r-1 \right) \tau _{n-1}\left( j,r-1\ \right)
\tau _{n}^{-1} \left( j,r-1\ \right) -
$$
$$
\tau_{n+2}\left(
j,r-1\ \right)\tau _{n+1}^{-1}\left( j,r -1\right)\tau _{n}\left(
j,r\ \right)  \tau _{n+1}^{-1}\left( j,r\ \right) +
$$
\begin{eqnarray} \tau _{n+1}\left( j-1,r\ \ \right)) \tau
_{n}^{-1}\left( j-1,r \right) \tau _{n}\left( j,r\ \right) \tau
_{n+1}^{-1}\left( j,r\ \right) = 0 \label{nH}
\end{eqnarray}

In the scalar case the system reduces to the Hirota bilinear
equation \cite{Mat}
\begin{eqnarray}\label{BH}
\tau _{n}\left( j+1,r\ \right) \tau _{n}\left( j,r+1\ \right) -\tau
_{n}\left( j,r\ \right) \tau _{n}\left( j+1,r\ +1\right) \ + \tau
_{n+1}\left( j+1,r\ \ \right) \tau _{n-1}\left( j,r+1\ \right) \ =0.
\end{eqnarray}

Using substitutions $\left( \ref{SUB}\right) $ and all DT
formalism we might give in a nonabelian form that is important in
connection with the applications in the theory of a quantum
transfer matrices for fusion rules $\ $\cite{Kun} and the theory
of quantum correlation functions \cite{Au}.

Let us return back to the DT theory. The equations $\left(
\ref{LT1}\right) ,\left( \ref{LT2}\right) $ are jointly covariant.
Hence the theory of solution of the systems
 $\left( \ref{HB2}\right) $  or $\left( \ref{nH}\right) $ is based on the
symmetry of the systems that is generated by the joint covariance
of the equations $\left( \ref{LH1}\right) ,\left( \ref{LH2}\right)
$ with respect to the one of transformations $\left(
\ref{DT}\right) ,$ namely to the

\bigskip
\begin{equation}
\ \psi^{-}(j,r)=\psi - \sigma ^{-}T^{-1}f,\;\sigma ^{-}=\varphi
\left( T^{-1}\varphi \right) ^{-1}.
\end{equation}

The form of the both linear equations $\left( \ref{LH1}\right) ,\left( \ref
{LH2}\right) $ is obvious reductions of the equations $\left( \ref{LT1}%
\right) ,\left( \ref{LT2}\right) $ with $V_{1}=1,V_{0}=v;
U_{0}=1,U_{-1}=u.$ We show further some details in the proof of
the covariance theorem because it shows the important features in
the procedure of the chain equation derivation. Let us start, say,
from $\left( \ref{LH2}\right) .$ The
conditions of the covariance are obtained as the coefficients by $%
\psi,T^{-1}\psi,T^{-2}\psi.$ The first one is valid automatically,

\begin{equation}
u^{-}=u - \sigma ^{-}\left( r-1\right) +\sigma ^{-}\left( r\right)
\label{DT1}
\end{equation}

\begin{equation}
u^{-}T^{-1}\sigma ^{-}\left( r\right) =\sigma ^{-}\left( r-1\right) u\
\label{DT2}
\end{equation}
Excluding the transformed potential\ $\ $\ $u^{-}$one arrives to the
equation that links the potential and the function $\sigma ^{-}\left(
r\right) $

\begin{equation}
\ u\ T^{-1}\sigma ^{-}\left( r\right) -\sigma ^{-}\left( r-1\right) u=\left(
\sigma ^{-}\left( r-1\right) -\sigma ^{-}\left( r\right) \right)
T^{-1}\sigma ^{-}\left( r\right) .  \label{Min}
\end{equation}
Note that the expressions in the equalities \eqref{DT1},
\eqref{DT2}, \eqref{Min} are still general (nonabelian) and may be
used in simplest (but definitely rich) closures (e.g.
$\sigma_{n+1}^- = \sigma_{n}^-$). The generic   is outlined in the
end of the sec. 3. For some Lie algebraic approach to the
nonabelian chains look also in \cite{Le}.

In the scalar (abelian) case one can easily solve the equation
\eqref{Min} with respect to the potential

\begin{equation}
\ u\ =\frac{\left( \sigma ^{-}\left( r-1\right) -\sigma ^{-}\left( r\right)
\right) T^{-1}\sigma ^{-}\left( r\right) }{T^{-1}\sigma ^{-}\left( r\right)
-\sigma ^{-}\left( r-1\right) }.  \label{Mia}
\end{equation}

Supplying the entries of the
equation $\left( \ref{Mia}\right) $ by the index N (iteration number) and
substituting into the equations $\left( \ref{DT1}\right) ,\left( \ref{DT2}%
\right) $ we obtain two equivalent forms of the chain equations. For example,
\begin{equation*}
u_{N+1}=u_{N}-\sigma _{N}^{-}\left( r-1\right) +\sigma _{N}^{-}\left(
r\right)
\end{equation*}
yields

\begin{equation}
\ \frac{\left( \sigma _{N+1}^{-}\left( r-1\right) -\sigma _{N+1}^{-}\left(
r\right) \right) T^{-1}\sigma _{N+1}^{-}\left( r\right) }{T^{-1}\sigma
_{N+1}^{-}\left( r\right) -\sigma _{N+1}^{-}\left( r-1\right) }=\frac{\left(
\sigma _{N}^{-}\left( r-1\right) -\sigma _{N}^{-}\left( r\right) \right) \ \
\sigma _{N}^{-}\left( r-1\right) \ }{T^{-1}\sigma _{N}^{-}\left( r\right)
-\sigma _{N}^{-}\left( r-1\right) }  \label{Ch}
\end{equation}

The chain equation $\left( \ref{Ch}\right) $ generates the chain equation
for the specific case of the system $\left( \ref{HB2} \right) $ by the choice $%
x\rightarrow n,Tf_{n\ }\left( j,r\ \right) =f_{n-1\ }\left( j,r\
\right) .$ A solution of the resulting chain equation generates
the solution of the system $\left( \ref{BH}\right) $ by use of the
connection formula $\left( \ref
{Mia}\right) $ and the corresponding formula for $v$. The transition to the $%
\tau _{n}\left( j,r\ \right) $ - functions is made by $\left( \ref{SUB}%
\right).$

\subsection{On solution of the chain equation}

Let us denote

\begin{equation*}
s_{N}=\frac{\left( \sigma _{N}^{-}\left( r-1\right) -\sigma _{N}^{-}\left(
r\right) \right) \ \ \ }{T^{-1}\sigma _{N}^{-}\left( r\right) -\sigma
_{N}^{-}\left( r-1\right) },
\end{equation*}

then the dressing chain equation $\left( \ref{Ch}\right) $ reads

\begin{equation}
s_{N+1}=s_{N}\frac{\ \sigma _{N}^{-}\left( r-1\right) }{T^{-1}\sigma
_{N+1}^{-}\left( r\right) }.  \label{Ch1}
\end{equation}

Iterating this recurrence yields

\begin{equation}
s_{N+q}=s_{N}\frac{\ \prod_{s=0}^{q-1}\ \sigma _{N+s}^{-}\left( r-1\right) }{%
\prod_{s=1}^{q}T^{-1}\sigma _{N+s}^{-}\left( r\right) }.  \label{sol}
\end{equation}

In analogy with continuous case let us consider the periodic closure of the
chain $\left( \ref{Ch}\right) $, starting from the simplest case q=0, that
means $\sigma _{N+1}=\sigma _{N}=\sigma .$ As$\ \ s_{N+1}=s_{N},$

\begin{equation*}
T\sigma _{\ }^{-}\left( r-1\right) =\sigma _{\ }^{-}\left( r\right) .
\end{equation*}

It means $\ \sigma _{\ }^{-}\left( r+p\right) =T^{p}\sigma _{\ }^{-}\left(
r\right) .$ If a boundary condition in the point r=0 is given, then

\begin{equation}
\sigma _{\ }^{-}\left( p\right) =T^{p}\sigma _{\ }^{-}\left( 0\right) .
\label{sig}
\end{equation}

The equation for $\varphi (p,x)$ then looks

\begin{equation*}
T\varphi (p,x)=T\left( \sigma (p,x)\right) \varphi (p,x).
\end{equation*}

The solution depends on the choice of T; e.g. if $T\varphi (p,x)=\varphi
(p,x+\delta )$, then

\begin{equation*}
\varphi (p,x)=\exp \left( Ax\right)
\end{equation*}

with A from

\begin{equation*}
A\delta =\ln \left[ T^{p+1}\sigma _{\ }^{-}\left( 0\right) \right]
\end{equation*}

\section{Nahm equation}

Considering the following example we change a bit the DT formulas,
showing the alternative version, similar to \cite{Sal}. We would
stress, however, that the formulas from the first section give the
equivalent result. Some  generalization we need within the
reduction constraints - we need` additional (gauge) transformation
denoted by $g$. This is expressed by the following
\begin{theorem}
The equation
\begin{equation}\label{eq}
\psi_y = u T\psi + v\psi + w T^{-1}\psi
\end{equation}
is covariant with respect to DT
\begin{equation}\label{DT1}
 \psi[1]=g(T-\sigma)\psi^{-1},
\end{equation}
Where $\sigma=(T\phi)/\phi,\qquad  \phi$ is a solution of the same
equation (\ref{eq}) and $g$ is an invertible element of the ring
to be defined later. The transforms of the equation coefficients
are
\begin{eqnarray}
u[1]=gT(u)[T(g)]^{-1},\\
v[1]=gT(v))g^{-1}-g\sigma ug^{-1}+gT(u)T(\sigma)g^{-1} + g_yg^{-1},\\
w[1]=g\sigma w [T^{-1}(g\sigma)]^{-1}. \label{pot}
 \end{eqnarray}
\end{theorem}

\begin{proof}
 The substitution of \eqref{DT1} into the transformed equation \eqref{eq} give four
equations assuming $T^n\psi$ are independent. Three of them yield transformed potentials
 \ref{pot}.
The fourth equation after use of the transforms takes the form
\begin{equation}\label{sy}
   \sigma_y = \sigma F-(TF)\sigma
\end{equation}
where
$$
F=u\sigma+v+w[T^{-1}(\sigma)]^{-1}.
$$
one can check the condition by direct substitution of the $\sigma$ definition and use
of the equation
for $\phi$.
\end{proof}

{\bf Remark} The theorem 2 is naturally valid for the spectral problem
\begin{equation}\label{sp}
 \lambda\psi=uT\psi+v\psi+wT^{-1}\psi
\end{equation}
with the only correction: the last term for the transform $v[1]$ is absent.
The equation goes to the "Riccati equation" analog for the function $\sigma$
 \begin{equation}\label{sp1}
    \mu = u\sigma+v+w[T^{-1}(\sigma)]^{-1},
\end{equation}
 Note that the plugging of the element
$\sigma=(T\phi)/\phi $ into the equation (\ref{sp1}) transforms it
to the spectral problem for $\phi$ (\ref{sp}) (with the spectral parameter $\mu$).

The Nahm equations could be written in the Lax representation using the spectral equation
(\ref{sp})
and the evolution equation
\begin{equation}\label{ev}
 \psi_y=(q+pT)\psi,
\end{equation}
The covariance of this equation with respect to the DT (\eqref{DT1}) may be established
similar to
the theorem 1 with account of the evolution for the function $\sigma(y)$
\begin{equation}\label{evs}
  \sigma_y =  T(q)\sigma-\sigma p\sigma+T(p)T(\sigma)\sigma-\sigma q=0,
\end{equation}
that proves the following transformation formulas for the coefficients in (\eqref{ev}).
\begin{equation}\label{p}
  p[1]= gT(p)[T(g)]^{-1},
\end{equation}
\begin{equation}\label{q}
 q[1] =g[T(q) - \sigma p + T(p)T(\sigma)]g^{-1}  + g_y g^{-1},
\end{equation}
The joint covariance principle \cite{L} define the connection of
potentials $p,q$ with $u,v$

\begin{equation}\label{pquv}
p=u+\beta I, \qquad q = v/2.
\end{equation}

Hence the DT-covariance means integrability of the compatibility
condition of (\ref{ev}),(\ref{eq})
$$
u_y = \frac{1}{2}(uT(v) - vu) + \beta(T(v)-v),
$$
$$
v_y = uT(w)-wT^{-1}u +\beta(T(w)-w),
$$
$$
w_y = \frac{1}{2}vw - wT^{}-1(v).
$$
One more possible specification is the use of periodic potentials
in the problem \ref{ev}) with the evolution (\ref{eq} with account
of the connections \eqref{pquv} that results in commutators in
r,h,s of the equations. Some linear transformations and rescaling
\begin{equation}\label{u}
  u = \alpha (-\imath  \varphi_1/2  - \varphi_3),
\end{equation}
\begin{equation}\label{v}
 v = \varphi_3,
\end{equation}
\begin{equation}\label{w}
w = \alpha^{-1}(-\imath  \varphi_1/2  + \varphi_3) ,
\end{equation}
produce Nahm equations (for the periodic functions $T\varphi_i =
\varphi_i$, that by the way does not mean a periodicity of
solutions of the Lax pair $\psi, \phi$ and the corresponding
$\sigma=(T\phi)/\phi)$
\begin{equation}
   \varphi_{iy} = \imath\epsilon_{ikl}[\varphi_k,\varphi_l].\label{Nahm}
\end{equation}
The numbers $\alpha, \beta $ are free parameters. This system is
covariant with respect to the combined DT-gauge transformations if
the gauge transformation $g = \exp G$ is chosen as follows.
\begin{equation}\label{G}
G_y = \alpha [(\varphi_3+\varphi_1/2)T(\sigma) -
\sigma(\varphi_3+\varphi_1/2)]/2,
\end{equation}
Finally, the following theorem may be formulated.

{\bf Theorem} The system \eqref{Nahm} is invariant with respect to the transformations
\begin{eqnarray} \label{DNahm}
 \varphi_1[1] = g[(\varphi_1/2 - \imath \varphi_3)T(g)^{-1}+
 \sigma(\varphi_1/2+\imath\varphi_3)[T^{-1}(g\sigma)]^{-1}],\\
 \varphi_2[1] = g[\varphi_2 + \alpha(\imath\sigma\varphi_1/2-
 \imath\varphi_1T(\sigma)/2 +\sigma\varphi_3 - T(\varphi_3\sigma))]g^{-1},\\
 \varphi_3[1] = g[(-\imath \varphi_1 /2-\varphi_3)T(g)^{-1}+
 \sigma(-\imath\varphi_1/2 + \varphi_3)[T^{-1}(g\sigma)]^{-1}]
 \end{eqnarray}
 with the function $g = \exp{G}$, where G is obtained integrating \eqref{G}.

{\bf Remark} Similar statement may be formulated for the discrete
version (see \cite{Mu}) of the Nahm system \eqref{Nahm}  as it may
be seen from the previous section.

 {\bf Conclusion}
The productivity of the method is obviously dependent on the
possibilities to solve chain equations. The direct applications of
DT formulas could be effective as it is well-known \cite{MS}. The
approaches to the direct solution of the chain equations is
formulated in \cite{VS} , \cite{Le} and refs therein.

{\bf Acknowledgement} This is an acknowledgement to V. Matveev, M.
Salle, M. Czachor and N. Ustinov for discussions, to  organizers
of the SIDE4 conference in Tokyo for the outstanding hospitality
and support, to A. Nakamula for the valuable information  about
Nahm model.
  The work was supported by the KBN Grant No. 2~P03B~163~15.

\end{document}